# Changed reference models in proportionality analysis.

Enrique Ordaz Romay[1]

*Facultad de Ciencias Físicas, Universidad Complutense de Madrid*

## Abstract.

In the statistical analysis of objects, samples and populations with quantitative variables, in many occasions we are interested in knowing the proportions that exist between the different variables from a same object; if these proportions have relations of normality among them, that is to say, if in the sample are these same proportions, and finally if the deviations of concrete physical or statistical normality have meaning.

The present study suggests the strategy of to use reference models for the analysis of the proportionality, using like example the studies of human proportionality from the method Phantom of Ross and Wilson and its extension towards scalable models.

## Introduction.

Let us suppose that we have a population formed by defined objects by means of quantitative variables. In order to make a study of such variables and their interrelations we must act of the following form:

1. To determine a representative sample of the population.
2. To make the measures and register the variables for each object of the sample.
3. To carry out the statistical tests and contrasts applicable to the values obtained during the measures of the sample objects.

---

[1] eorgazro@cofis.es

One of the first questions that arise in any statistical study of populations is: What relation exists between the different variables measured for a same object?. That is to say, How the intra-individual proportions are related to the proportions of the variable samples?.

In order to respond to these questions in the present study we raised the strategy to use reference models for the analysis of the proportionality. That is to say, ideal models of objects that are used like standard of measurement in proportionality. Some times, will be models formed by means of statistical averages, other times convenient or useful mathematical models to even facilitate the calculations and in some cases, the models will be established like canons of some specific quality.

The methods of analyses used in our study will be applied to the studies of human proportionality through the Phantom method of Ross and Wilson, extending their initial utility like method to quantify the proportions, towards a more complete statistical interpretation and an extension towards scalable strategies of the Phantom model.

## The proportionality concept.

The proportionality is the mathematical property according to which two objects *A* and *B*, each one with two quantitative variables: *variable*$_1$ y *variable*$_2$ (with equal dimensions); they fulfill the following property:

$$\frac{variable_1(A)}{variable_1(B)} = \frac{variable_2(A)}{variable_2(B)} \qquad (1)$$

If the equation (1) is fulfilled it says that the variables: *variable*$_1$ y *variable*$_2$ , in the objects *A* and *B*, are in proportion.

Nevertheless, the expression (1) is only valid when the dimensions of *variable*$_1$ and *variable*$_2$ are the same ones. When the dimensions of these two properties are $n_1$ and $n_2$ respectively, the equation (1) becomes general of the form:

$$\left(\frac{variable_1(A)}{variable_1(B)}\right)^{1/n_1} = \left(\frac{variable_2(A)}{variable_2(B)}\right)^{1/n_2} \qquad (2)$$

Let us suppose that the relation (2) was not fulfilled because the objects *A* and *B* were not in proportion, but we were interested in calculating what Δ*x* value would be necessary to add to *variable*$_1$(*A*) in order to fulfill the relation (2). Let us call *n* to the ratio between two dimensions $n = \frac{n_1}{n_2}$. In these conditions it is verified:

$$\left(\frac{variable_1(A) + \Delta x}{variable_1(B)}\right) = \left(\frac{variable_2(A)}{variable_2(B)}\right)^n \Rightarrow \qquad (3)$$

$$\Delta x = variable_1(B)\left(\frac{variable_2(A)}{variable_2(B)}\right)^n - variable_1(A)$$

The value of Δ*x* can be interpreted like the difference or existing proportional distance between the objects *A* and *B* respect to the variables *variable*$_1$ and *variable*$_2$. To variable *variable*$_2$ is called proportionality base.

Usually, in the proportionality studies, the comparison is not made between two objects nobodies, but between an object *B* and another *P* that we used like reference or standard.

Let us suppose, as it is used normally, that the *P* standard to that we will call reference model, is not a real object, but the result of the statistical distribution of a representative sample of a population in which it is including object *B*. In these conditions, if the distribution of the variables of the sample is a normal distribution, the reference one will be defined using a table that contains the values of the averages and standard deviations of the different variables from the sample. That is to say, the reference model *P* will be defined in the form of table 1

| Variable | Average | Standard deviation |
|---|---|---|
| *Variable₁* | *Average₁* | *StdDev₁* |
| *Variable₂* | *Average₂* | *StdDev₂* |
| *Variable₃* | *Average₃* | *StdDev₃* |
| … | … | … |
| *Variableₘ* | *Averageₘ* | *StdDevₘ* |

Table 1. Prototype of table for reference P.

For example: In the studies of human proportionality, Ross and Wilson in 1974 defined a proportionality standard to which they called "Phantom" model so that it served like reference. The calculation of the values corresponding to the kinanthropometric variables of the Phantom model was based on extensive data bases of general population. The perimeters were obtained from the data base of Wilmore and Behnke in 1969 and 1970, the skinfolds from a data base not published of Yuhasz (Carter, 1996) and the rest of measures was obtained by Garret and the Kennedys in 1971. The result of these works lead until the table of the Phantom reference model.

| Variable | Average | Standard deviation |
|---|---|---|
| *Height* | 170.18 *cm* | 6.29 *cm* |
| *Acromial height* | 139.37 *cm* | 5.45 *cm* |
| *Radial height* | 107.25 *cm* | 5.37 *cm* |
| … | … | … |
| *Weight* | 64.58 *kg* | 8.60 *kg* |

Table 2. Fragment of the table of the human model "Phantom".

When we want compute the distance between an object *i* and one sample, *Z* score like the transformation rule is defined that allows compute the absolute distance between both. Its calculation is made as the difference between value $x_i$ and the average of the sample, divided between standard deviation *s* of the sample:

$$Z_{muestra}(i) = \frac{x_i - \bar{x}}{s} \tag{4}$$

The value of $x_i - \bar{x}$ is equivalent to the proportionality equation (3), that indicates the proportional distance between two variables. Replacing $x_i - \bar{x}$ by $\Delta x$ in the expression (4) we have left:

$$Z_{variable_1,variable_2;P}(i) = \frac{variable_1(i)\left(\frac{variable_2(P)}{variable_2(i)}\right)^n - variable_1(P)}{s_{variable_1}(P)} \tag{5}$$

This is the equation that is used for the calculation of the proportional distance between an individual *i* and a reference *P* respect to the variable $variable_1$ being $variable_2$ the proportionality base. Being the proportionality base, in many occasions, inherent part of the calculation method of the proportionality, gets used to omitting this subscript.

Returning to the example of the model of human proportionality "Phantom", the method that traditionally is applied for the calculation of *Z* scores uses as proportionality base to the height variable, that is to say, $variable_2 = height$. On the other hand, since the height is a linear variable, $n_2 = 1$ and therefore $n = n_1$. That is to say, the explaining *n* agrees with the dimension of $variable_1$. Thus, calling *Ph* to the reference model "Phantom", the equation (5) is transformed into:

$$Z_{variable;Ph}(i) = \frac{variable(i)\left(\frac{height(Ph)}{height(i)}\right)^n - variable(Ph)}{s_{variable}(Ph)} \tag{6}$$

That it is the equation used in kinanthropometric, in the analysis of the proportionality, for the method "Phantom" of Ross and Wilson.

## The statistical interpretation of *Z* score in proportionality.

From the statistic and under the assumption of normality of the referet model P, we found the interpretation of *Z* scores in proportionality. Let us suppose, at a first moment, that $variable_2(i) = variable_2(P)$. [2] In this assumption we can rewrite the equation (5) of $Z_{variable;P}(i)$ in the form:

$$Z_{variable;P}(i) \cdot S_{variable}(P) = variable(i) - variable(P)$$

That is to say, Z score in proportionality indicates, proportionally, the number of times that is greater or smaller the standard deviation than the difference between the value of the variable in object *i* and the reference *P*

Then, under the assumption that the reference *P* is distributed of normal way (Gaussian), the value of Z indicates the position that corresponds to the object within the distribution function. This measurement of position, using a table of normal distribution, can be transformed easily into a value of percentile of the study object within the sample. Table 3 shows some examples of percentiles near zero.

| **Z** | -3 | -2 | -1 | -0.5 | -0,2 | 0 | 0,2 | 0.5 | 1 | 2 | 3 |
|---|---|---|---|---|---|---|---|---|---|---|---|
| **Percentile** | 0,1 | 2,3 | 15,9 | 30,9 | 42,1 | 50 | 57,9 | 69,1 | 84,1 | 97,7 | 99,9 |

Table 3 Values of the percentiles based on the value of Z score.

In the example of the model of human proportionality "Phantom", nevertheless, this interpretation is only a first approach to the complete interpretation of Z score because the own model creates several problems.

1. By construction, the Phantom model, theoretically is not normal, because for the calculation of the values of the variables was used a heterogeneous group formed by men and women of diverse ages (all adults) and ethnic groups.

2. Another factor of not-normality is that the values of the measured variables are always positive and therefore, exists a minimum value for each Z score:

---

[2] From this section we will consider that *variable₂* like proportionality base is something inherent to the calculation method of scores Z of proportionality, as we did in the equation of the model of human proportionality "Phantom" in which it is implied, for future references, that *variable₂* = *height* and is omitted like subscript of Z.

$$Z_{variable;Ph}(minimum) = -\frac{variable(Ph)}{s_{variable}(Ph)}$$

Consequently, the interpretation of $Z$ score of the Phantom model like a percentile, cannot be a complete interpretation, but only a first approach that will have later to be clarified based on characteristics like: sex, age, diet, physical activity or ethnic group of the subject of study $i$.

## Causes for the change of the reference model in the proportionality analysis.

One of the most important problems in the proportionality studies is the change from a one reference model to another one. The reason that these transformations suppose a problem is because the results of $Z$ scores in two different reference models will give like result values of different $Z$ scores and therefore different (although compatible) interpretations.

Indeed, the problem resides in the search of the "compatibility way" of the made interpretations using different reference models. Once found this way, we will be able to go from a one reference model to another one according to the interests.

Another reason to look for this "compatibility way" is that some reference models do not come from normal samples (as it happened in Phantom model). Nevertheless, other reference models can give a precise statistical interpretation to $Z$ scores when coming from normal samples.

A universal reference model for proportionality supposes an advantage because all the studies will make reference to the same standard and the results and conclusions will be compatible. On the other hand, the interpretation of $Z$ scores in universal models, which, in general, are heterogeneous, will be less precise and will require to clarify them.

On the other hand, the particular reference models of homogenous groups will facilitate values of very precise *Z* scores in their interpretation although their results and conclusions will not be general without a compatibility way.

In the models of human proportionality, in addition to the model "Phantom" of Ross and Wilson we can find reference models based on the qualities of the different samples studied in the different scientific works. Three types of particular reference models in human proportionality, each one divided by sex exist basically:

- Regional Models: they are those that gather the own characteristics of a geographic zone which, due to the genetics, climate, diet and activity, marks in its population a characteristic proportionality. They are classified by geographic regions and they are denoted like *Reg*(*zone*, *sex*).

- Sport Prototypes: they correspond to the own values of each sport. Thus, the prototype of athlete of weightlifting will be very different from the prototype of long-distance runner. They are classified by sport modalities and they are written like *Spt*(*model*, *sex*).

- Ontogenic Models: they correspond to the changes in the proportionality due to the age. They are classified by age and they are denoted by *Ont*(*age*, *sex*).

The compatibility way between the different reference models would allow to go from any model (regional, ontogenic or sport) to the Phantom model and vice versa, so that the results and conclusions of the different studies can be generalized.

# Method for the change of the reference model in the proportionality analysis.

Let us suppose that we have two reference models *P* and *P'*. Immediately, than to replace in the equation (5) it is observed that, under the condition:

$$\frac{variable(P)}{s_{variable}(P)} \approx \frac{variable(P')}{s_{variable}(P')} \qquad (7)$$

verifies to the equation: $\qquad Z_{variable;P}(i) \approx Z_{variable;P}(P') + Z_{variable;P'}(i) \qquad (8)$

The expression (8) is the compatibility way looked between $Z$ scores from two different reference models. That is to say, under the adapted relations between the reference models $P$ y $P'$ given by the expression (7), proportionality $Z$ scores respect to a one reference can be break in sum of $Z$ scores respect to another one reference plus an amount that calculates like $Z_{variable;P}(P')$ in the equation (5). The interpretation of $Z_{variable;P}(P')$ is the one of $Z$ score respect to variable corresponding to the averages of the reference model $P'$ respect to the reference model $P$.

That is to say, if $P$ represents a universal reference model (general and heterogenous) and $P'$ represents a particular reference model (special and homogenous) $Z_{variable;P}(P')$ indicates the proportional variations due to the particularities of $P'$ respect to $P$.

The equation (8) can be understood from a dispersion graph in which, $Z$ scores are represented in the x-axis, whereas the variables are written in the y-axis. This representation can be seen in graph 1.

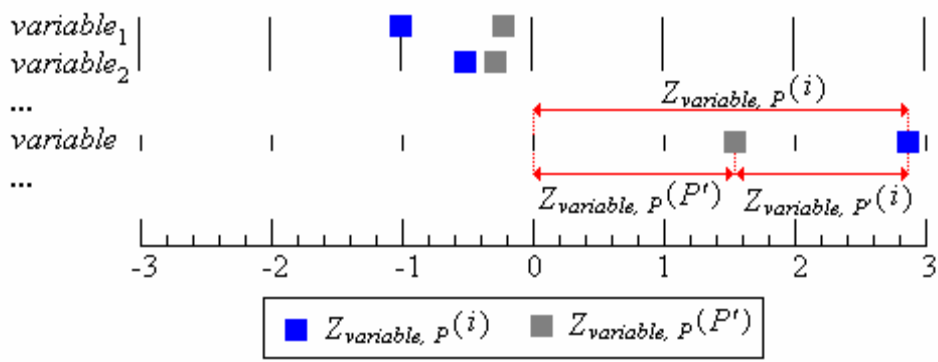

Graph 1. Dispersion graph for the representation of proportionality $Z$ scores. Graphical representation of the equation (8): Break of $Z$ scores for two reference models $P$ and $P'$

In the case of the model of human proportionality: let us take an individual *i* with *age* = 7 years y *sex* = *female* with the variable *triceps skin-fold* = 8,7 *cm* and the *height* = 115,6 *cm*. Looking for in the table of the Phantom model, for this variable we found:

- *triceps skin-fold* $(Ph)$ = 139,37 *cm*
- $s_{\text{triceps skin-fold}}(Ph)$ = 5,45 *cm*.

A simple substitution of these values in the equation (6) gives by result an score: $Z_{\text{triceps skin-fold; Ph}}(i) = -0{,}6001$, that is to say, it folds of triceps is in the 27,42 percentile.

At first sight, the physical interpretation of this result could be that this individual one shows proportionally far below *triceps skin-folds* respect to the Phantom model. Nevertheless, this result throws a doubt: Are the values obtained for individual *i* in proportion with a homogenous sample of individuals of the same characteristics?.

In order to respond to this question we consulted an ontogenic model. Let us suppose that we found the values:
- *Triceps skin-fold*(*Ont*(7,*female*)) = 8,95 *cm*; $s_{\text{triceps skin-fold}}(Ont(7,female))$ = 2,51 *cm*.
- *Height*(*Ont*(7,*female*)) = 120,20 *cm* ; $s_{\text{height}}(Ont(7,female))$ = 4,44 *cm*.

A substitution of these values in the equation (5) leads a: $Z_{\text{triceps skin-fold; Ont}(7,femenino)}(i) = 0{,}0383$ or what is the same a 51,52 percentile. That is to say, individual *i* shows a *triceps skin-fold* proportionally very similar to the average value for its age, although with respect to the reference model "Phantom" the result was smaller.

When we calculated $Z_{\text{triceps skin-fold;Ph}}(Ont(7,female))$ score can be found the reason of this discrepancy: its result is –0,6316, or what is the same, we found a 26,38 percentile. That is to say, the reference model *Ont*(7, *female*), in triceps skin-fold shows a disproportion inherent or characteristic of the own individuals of the sample.

This event is known as alometric growth (each part of the anatomy grows at a different speed) and it is what produces these differences between the values of different *Z* scores. Nevertheless, in human proportionality we are more interested in knowing *Z* score of *i* respect to the *Ont*(7, *female*) that respect to the Phantom model, because in the ontogenic models the alometrics changes are already assumed.

On the other hand, like *Ont*(7,*female*) verified (7) equation, and :

$$Z_{triceps\ skin\text{-}fold;Ph}(Ont(7, female)) + Z_{triceps\ skin\text{-}fold;Ont(7,femenino)}(i) = -0{,}5932$$

is approximately equal to $Z_{triceps\ skin\text{-}fold;\ Ph}(i) = -0{,}6001$ with a relative error 1,13%, this example also serves to illustrate the equation (8).

This method for proportionality analysis by means of change from a one reference model to another one in human proportionality was called "scalable model" and was proposed in 2005 in the XXII congress of Sociedad Anatómica Española in Murcia (Spain).

## Conclusion

When we work in proportionality, with reference models, we are always between two commitments: What reference model must we use?: universal or particular.

- We will look for a universal reference model that it serves in a great population so that all the scientific works can use it and thus to be able to share the results and conclusions of each study. In return, the sample of reference universal model will be heterogeneous and its distribution will be not-normal, making difficult the interpretation of *Z* scores.
- On the contrary, particular reference models will work with specific homogenous samples for the differentiating characteristics of the population. Calculated *Z* scores using these models will be more easily interpretable and will contribute much information. In return, the results and conclusions obtained, using particular reference models our studies will not be directly generalizable to another studies.

A good solution for this dilemma consists of publishing, in addition to the tables of universal reference models, also the tables of special reference models, so that, using the equation (8) one score can be transformed into another.

Distinguishing to the case of the human proportionality, it would be necessary to obtain and to publish tables of ontogenic, regional and sport models referring to allow to the study of the individual proportionality of the respect samples with their same characteristics and their generalization to universal model Phantom through the equation (8) to facilitate the interchange of results and conclusions between the different studies.